\documentclass[12pt]{article}

\newcommand{\be}{\mbox{\boldmath$e$}}
\newcommand{\bp}{\mbox{\boldmath$P$}}

\newcommand{\bfq}{\mbox{\boldmath$Q$}}
\newcommand{\br}{\mbox{\boldmath$r$}}

\begin{document}

\title{Partial cancellation of correlations in nonrelativistic
high energy asymptotics of photoeffect}

\author{E. G. Drukarev$^1$ and  R. H. Pratt$^2$\\
$^1$  Petersburg Nuclear Physics Institute\\
Gatchina,  St. Petersburg 188300, Russia\\
$^2$ Department of Physics and Astronomy, \\ University of Pittsburgh, Pittsburgh, PA 15260 USA}
\date{}
\maketitle

\begin{abstract}
We investigate the total effect of correlations on photoionization
of atomic states with  nonzero orbital momentum, in the
nonrelativistic high energy asymptotic limit, considering the
exclusive case of the dominant final state of an initial neutral
atom. We find that the substantial cancellation of the dominant
intra-shell correlations, which had been reported earlier, can be
understood utilizing the closure properties satisfied by the
eigenfunctions of the nonrelativistic Hamiltonian. Considering the
sum of correlations with  all states, occupied or not, we show
that complete sum is equal to the contribution of the high energy
part of the continuum. Consequently there is a total cancellation
between the contributions of the bound states and the low energy
part of the continuum states. This means that the real
correlations in the physical atom, due to the sum rule over the
occupied states, can be also obtained as the negative of the total
contribution of low energy bound and continuum unoccupied states.
We calculate this in the framework of the quantum defect model. As
we would expect, the results are close to those obtained earlier
in particular cases by direct summation over the occupied states.
However this approach also allows us to see that the dominant
intrashell correlation is going to be cancelled. We can also
obtain some
 limits on
the correlation effects by considering calculations with the
screened Coulomb functions.  The role of correlations in the
exclusive photoionization processes is discussed, also the
modification of correlations in the case of atomic ions.

\end{abstract}

\section{Introduction}

In this paper we study the total effect of the final state
correlations on the amplitudes and cross sections of
photoionization of atomic states with nonzero values of orbital
momenta in the nonrelativistic high energy asymptotic limit. It is
known that in this situation correlation effects still persist
even in the high energy limit. We show that the sum of such
correlations shows a tendency to cancel. The experimental data for
photoionization of $p$ states of external shells of neon and argon
by photons of the energies of about 1 keV \cite{1}, \cite{2} can
not be interpreted in the framework of the Independent Particle
Approximation (IPA). In IPA the $p$ electron is ionized by direct
interaction with the photon. The authors of \cite {1} suggested a
mechanism of IPA breaking by the final state electron
interactions. The photon ionizes rather the $s$ electron of the
same subshell. In a next step the outgoing electron moves the $p$
electron to fill the hole in the $s$ state created by the photon.
The final state correlations have been studied later in
\cite{3}-\cite{5}, with the initial state IPA breaking effects
being included in \cite{4},where they were shown to be small
except some special cases. It was understood that at still larger
energies, much greater than all the binding energies of the atom,
there is a large cancellation between correlations with various
shells. Such cancellations were first found in the angular
distributions of photoionization \cite{3}, and later at the
amplitude level \cite{5}.

Since calculations involving cancellations require more precise knowledge
of wave functions for the description of the bound electrons,
we try to demonstrate the cancellations in another way.
We consider {\em asymptotic energies} of the outgoing electron,
$$
E\ \gg\ I,
$$
where $I$ is the ionization energy of the single-particle ground
state, and we seek to obtain the asymptotic amplitude. Thus we
assume $E$ to be much larger then all single-particle bound state
energies.  Our analysis is completely nonrelativistic, i.e, we
assume $E \ll m$ with $m$ for the electron mass (we employ the
system of units with $\hbar=c=1$). We consider only relatively
light atoms, with not very large values of the nuclear charge $Z$,
describing the bound electrons by nonrelativistic functions, with
corrections of the order $(\alpha Z)^2$ being neglected. We focus
on the case of $p$ electron photoionization.

Assuming that all initial electrons are moving in the same
self-consistent field, we show that these cancellations can be
understood utilizing the closure properties satisfied by the
eigenfunctions of the nonrelativistic Hamiltonian. Considering the
sum of correlations with  all states, occupied or not, we
demonstrate that complete sum is equal to the contribution of the
high energy part of the continuum. Thus there is total
cancellation between the contributions of the bound states and the
low energy part of the continuum states, for which we will give a
precise definition. Hence the sum of the correlations can be
expressed as the negative of the sum over low energy unoccupied
states.

We perform some explicit calculations for real atoms  by
calculating the contribution of  low energy unoccupied states,
using the quantum defect model combined with the Fermi-Segre
theorem, and making a rough estimate of the contribution of the
low energy part of the continuum. In our approach all such terms
are positive. Our results are close to those obtained by direct
summation over the occupied shells. These direct terms occur with
both signs,  in the cases where such summations were carried out
\cite{3,5}.

We have carried  out direct calculations here for the Coulomb case,
and we find certain limits on the correlation effects in this case.
Using perturbative treatment of the screening we show that the
magnitude of cancellations in the real atom is greater.

In our analysis we have used the perturbative approach to the
final state interactions of the electrons  developed in \cite{6}.
This approach was employed earlier for investigation of the IPA
breaking
 in photoionization \cite{4}. Inclusion of these effects removed
or strongly diminished the discrepancy between experimental data
and the IPA calculations. We shall use the expressions obtained in
\cite{4} throughout the paper.

In particular calculations we use the simplifying assumption that
overlap matrix elements between the orbitals of different
subshells in the initial state ion and the final state ion is
small and can be neglected. In this approximation the inclusive
cross section coincides with the exclusive one in which the state
of the spectator electrons does not change.
%In most of the paper we focus on inclusive processes, in which
%only one of the atomic electrons moves to another state. These
The latter cross sections correspond to the experiments
\cite{1,MN}. We show that the correlations considered in the paper
can also manifest themselves stronger in inclusive processes of
photoionization accompanied by excitation of external electrons.
In the case of atomic ions there will be less cancellation among
the correlations.

The paper is organized as follows. In Sec 2. we recall the main
equations for the perturbative treatment of the IPA breaking
effects in photoionization. In Sec.~3 we write the sum rules
provided by closure and show that they have the consequence that
there is  total cancellation between the sum over all the bound
states and the low energy continuum states. In Sec.~4 we obtain
the correlations of the occupied states, both directly (from
previous work \cite{5}), and as the negative of the sum over
unoccupied low energy states. In Sec.~5 we make explicit
calculations, using some simplified models. We show the results of
these approaches in Table~2, including also the cases of large $Z$
for illustrative purpose. For the direct calculations we
investigate the Coulomb case in Sec.~6, with the results given in
Table~3, and give some discussion of screening in an effective
charge approach.  In Sec.~7. we consider the role of correlation
in exclusive processes, and for atomic ions. We summarize in
Sec.~8. Some details of computations are presented in Appendices.

%Phys. Soc.  {\bf44}, 132 (1999).

%\bibitem{MN} O. Hemmers {\em et al.}, J. Phys. B~{\bf30}, L727 (1997).

%\bibitem{6} H. K. Tseng, and R. H. Pratt, Phys. Rev. A~{\bf3}, 100
%(1971).
%\bibitem{7} H. Bethe and E. E. Salpeter, {\em Quantum mechanics of One-
%and Two-Electron \\ Atoms} (Springer-Verlag, Berlin, 1957).
%\bibitem{8} F. Herman, and S. Skillman, {\em Atomic Structure Calculations}
%NJ, 1963.
%\bibitem{M} S. T. Manson, Phys. Rev. {\bf182}, 97 (1969).
%\bibitem{10} A. Kratzer and W. Franz, {\em``Transzendente Funktionen"},
%Leipzig, 1960.
%\bibitem{6} T. Kato, Come Pure Appl. Math. {\bf10}, 151 (1957).

\section{Perturbative treatment of IPA breaking effects}

We recall  the general points of our perturbative approach
\cite{6}, restricting ourself here to asymptotic analysis. We
shall use a simplifying assumption that the bound electrons are
described by single-particle wave functions. This is not a
necessary assumption, and the approach has been employed for the
case of correlated functions as well \cite{DT}, \cite{DA}.Consider
the asymptotic amplitudes $F_i$ for ionization of initial state
$i$ with quantum numbers $i=n,\ell,\ell_z$. The final state
interactions between the outgoing electron and the residual ion in
their lowest order in the the amplitude beyond the independent
particle approximation (IPA) can be expressed in terms of a linear
combination of the IPA asymptotic amplitudes $F_j^{0}$ for
ionization of all the other occupied atomic states $j$:
\begin{equation}
F_i\ = \ F_i^{0}+\sum_j F_j^0\Lambda_{j,i}\ ,
\end{equation}
with $\Lambda_{j,i}$ the matrix element for a transition from the
state $i$ to the state $j$, caused by the outgoing electron,
following photoionization of the state $j$.  If $j$ is a bound
$n's$ state (this will the most important case, since correlations
with higher $\ell$ states contribute beyond the asymptotics
\cite{4}), the asymptotics of the IPA amplitude can be written as
\cite{BS,A}
\begin{equation}
F^0_j\ =\ \langle\psi_f^0|\hat\gamma|\psi_j\rangle\ =\
\frac{(4\pi)^{1/2}\tau}{m} \frac{(\be \cdot \bp)}{P^4}N^r_{n's}\, ,
\end{equation}
with $\psi_f^0$ the plane wave approximation for the wave function
of the outgoing ejected electron, $N^r_{n's}$ the normalization
factor of the radial function of the $n's$ electron (
$\psi_{n's}(\br)=\psi^r_{n's}(r)/\sqrt{4\pi}$;
$N^r_{n's}=\psi^r_{n's}(0)$ , $\tau=m\alpha Z$, and
$\hat\gamma=-i(\be\cdot \mbox{\boldmath$\nabla$})/m$  is the
interaction operator between the photon with polarization vector
$\be$ and the electron.

The plane wave approximation
\begin{equation}
\psi_f^{(0)}(\br)\ =\ \exp[i(\bp\ \cdot \br)],
\end{equation}
with momentum of the outgoing electron $P\gg\tau,$ is appropriate
for Eq.~(2) in velocity form. This corresponds to normalization of
the continuum wave functions by the condition \cite{LL}
$$\int d^3r\psi^*_{\bp}(\br)\psi_{\bp'}(\br)=(2\pi)^3\delta(\bp-\bp')$$
The amplitude $F_j^0$ is evaluated in Appendix~A.

If the photon energy well exceeds the binding energies of the
bound states $i$ and $j$, the matrix elements $\Lambda_{j,i}$ can be
represented as
\begin{equation}
\Lambda_{j,i}\ =\ i\xi S_{j,i}\ ,
\end{equation}
with $\xi=m\alpha/P$ ($P$ the momentum of the outgoing electron)
the Sommerfeld parameter
of the final state interaction of the outgoing electron with
the residual ion. The matrix elements
\begin{equation}
S_{ji}\ =\ \langle j|\ln(r-z)|i\rangle\ ,
\end{equation}
(with $z$ the projection of the coordinate vector {\br} on the
direction of the momentum of the outgoing electron), obtained in
\cite{1}, describe the transfer of an electron from the states $i$
to fill the hole in the state $j$ of the positive ion with the
hole in $i$ state. One can write $\ln(r-z)=\ln r+\ln(1-t)$, with
$t=(\bp\cdot\br)/Pr$ and $\bp$ the momentum of the outgoing
electron. For states $i$ and $j$ with different angular momenta
the  matrix element (5) with $\ln r$ vanishes due to orthogonality
of the angular parts of the wave functions of the states $i$ and
$j$. Thus when the states $i$ and $j$ have different angular
momenta we can write
\begin{equation}
S_{j,i}\ =\ \langle j|\ln(1-t)|i\rangle\,.
\end{equation}
Only such states contribute to the asymptotic amplitude, when $i$
is not an $s$ state (which we will assume), since the dominant
asymptotic amplitudes $F_j^0$ require $j$ be an $s$ state.  Thus,
we shall consider only the states $j$ with quantum numbers
$\ell=\ell_z=0$, i. e. $j=n^{'},0,0$.  Taking the direction of the
outgoing electron momentum as the axis of quantization of angular
momentum, we find that correlations occur only for $i$ states with
$\ell_z=0$, since only these states are coupled by Eq.~(6).

We may therefore write
\begin{equation}
 F_i=F^{0}_i+i\xi\Sigma_j A_{j,i}\,,
\end{equation}
where
\begin{equation}
A_{j,i}\ =\ F^{0}_jS_{ji}\,,
\end{equation}
with $S_{j,i}$ given by Eq.~(6) for photon energies, well
exceeding the binding energies of $i$ and $j$ states (being
suppressed otherwise \cite{1,6}). For example, as shown in
\cite{1}, in ionization of the $2p$ electrons of neon by photons
with  energies of about  1 keV correlations with $2s$ electrons
are important, while those with $1s$ electrons are not; but by
10~keV correlations with $1s$ electrons become important. We will
write $A_{ji}$ as $A_j$ omitting the index $i$.

Now we restrict ourselves to the case of ionization of $p$ states
only, i. e. we consider the case $i=n,1,0$.  We use the standard
spectroscopic notation, e.g. the state with quantum numbers $n,0,0$ is
an $ns$ state, etc.  Then Eq.~(6) can be evaluated as
\begin{equation}
S_{ji}\ =\ -\frac{\sqrt3}2
\langle\phi^r_{n's}|\psi^r_{np}\rangle\,,
\end{equation}
with the first factor coming from the angular integration,
$\psi^r_{np}$ and $\phi^r_{n's}$ are respectively the radial wave
functions in the field of the atom and of the ion with the hole in
$np$ state, and
$$\langle \phi^r_{n's}|\psi^r_{np}\rangle\ =\int dr
r^2\phi^r_{n's}(r)\psi^r_{np}(r).$$

Note that since all the other electrons in initial and final
states belong to different Hamiltonians, there are nonzero overlap
integrals between orbitals of different subshells. This makes the
whole picture more complicated-see \cite{JC}. For example, in our
case there are other channels for ionization of $2p$ state. In one
of them the photon interacts directly with $2s$ electron, while
the $1s$ electron suffers shakeup into the hole in the $2s$ state
of the ion. The photoelectron pushes the $2p$ electron into $1s$
hole of the final state ion at the end of the story. The
contribution of this channel to the total amplitude of ionization
from the $2p$ state is thus $ F^{0}_{2s}\Lambda_{1s,
2p}\langle\phi^r_{2s}|\psi^r_{1s}\rangle$. If
$|\langle\phi^r_{2s}|\psi^r_{1s}\rangle| \ll 1$ we can neglect
this contribution with respect to the other terms on the right
hand side of Eq.(1).

We shall consider this very case, thus assuming that
$|\langle\phi^r_{n's}|\psi^r_{ns}\rangle| \ll 1$ for all $n\neq
n'$. Hence we find $1- |\langle\phi^r_{ns}|\psi^r_{ns}\rangle|^2
\ll 1$ for all $n$, and the inclusive cross section with the sum
over all possible states of the final ion coincides with the
exclusive cross section in which the spectator electrons do not
suffer transitions. In this approximation we must replace Eq.(9)
by
\begin{equation}
S_{ji} = -\frac{\sqrt 3}{2}\langle
\psi^r_{n's}|\psi^r_{np}\rangle.
\end{equation}

%The upper index $r$ denotes the radial parts of the bound state
%wave functions, and
%$$
%\langle \psi^r_{n's}|\psi^r_{np}\rangle\ =\int dr
%r^2\psi^r_{n's}(r)\psi^r_{np}(r).
%$$
Hence,
\begin{equation}
A_{n's}\ =\ -\frac{\sqrt3}2 F^0_{n's}\,\langle\psi^r_{n's}|
\psi^r_{np}\rangle\ ,
\end{equation}
with $F^{0}_{n's}$ the asymptotic IPA amplitude for ionization of
an $n's$ bound state. We shall omit the upper index IPA further.
For correlations inside the same shell
$$ A_{ns}
=-\frac{\sqrt3}{2}F^0_{ns}\langle\psi^r_{ns}|\psi^r_{np}\rangle.$$
For ionization of $2p$ states, i. e. $n=2$,
\begin{equation}
\langle \psi^r_{2s}|\psi^r_{2p}\rangle\ \approx\ -1\,,
\end{equation}
for all atoms. This matrix element, calculated with Hartree-Fock wave functions, is $-0.91$ for $Z=5$, with the value becoming closer
to $-1$ for larger $Z$. The Coulomb value is $-\sqrt 3/2$.Thus we can estimate that
\begin{equation}
A_{2s}\ =\ \frac{\sqrt3}2 F^0_{2s}\ .
\end{equation}

For  photon energies well exceeding the bounding energy of $L$
shell, but not of the $K$ shell, the correlation with the $2s$
electron dominates. Correlations with $1s$ electrons are small at
these energies \cite{1,4}.  Correlations with other $s$ electrons,
if there are any, are small This happens for two reasons. The
overlap matrix elements $|\langle \psi^r_{n's}|\psi^r_{2p}\rangle
|\ll  |\langle \psi^r_{2s}|\psi^r_{2p}\rangle|$, for $n'\neq 2$,
due to Eq.(12) and to the closure relation $\sum_{xs} |\langle
\psi^r_{xs}|\psi^r_{2p}\rangle\ |^2=1$ (with summation over the
states of both discrete and continuum spectra) Also the asymptotic
IPA amplitudes $F^0_{n's}$ drop with $n'$. Hence at such energies
the value (13) determines the scale of IPA breaking effects.

However, at larger energies, greatly exceeding the binding energy
of the $1s$ electrons, $A_{1s}$ becomes comparable to $A_{2s}$,
and there is a large cancellation between contributions of the $K$
shell and the other shells. Such cancellations were first found in
the angular distributions of photoionization \cite{3} and then
observed at the amplitude level \cite{5}. The calculations require
knowledge of rather  precise wave functions for the description of
the bound electrons.

For photon energies much greater than all the binding energies
 the asymptotic contribution of correlations to the amplitude is
\begin{equation}
\tilde T_{d}\ = \Sigma_{\tilde j}A_{\tilde j}=\Sigma_{\tilde ns}
A_{\tilde ns}\,,
\end{equation}
with the sum over all occupied $\tilde ns$ states, where
$A_{\tilde ns}$ is given by Eq.~(11), for all occupied states
$\tilde j$, where $A_{j}$ is given by Eq.~(8), namely
\begin{equation}
A_{\tilde j}\ =\ \langle\psi^0_f|\hat\gamma|\psi_{\tilde j}\rangle
\langle \phi_{\tilde j}|\ln r(1-t)|\psi_i\rangle \approx
\langle\psi^0_f|\hat\gamma|\psi_{\tilde j}\rangle \langle
\psi_{\tilde j}|\ln r(1-t)|\psi_i\rangle.
\end{equation}

There are only two active electrons in our analysis, while the
others are just ignored. This corresponds to calculation of the
amplitude for an inclusive process, since the sum over all
possible final states $\Phi_f$ of the spectator electrons in the
final state ion, described in the initial state atom by the
function $\Psi$
$$\sum_f\langle \Phi_{f}|\Psi\rangle^2=1$$.

% correlation
%he partial cancellation
%of the correlations means that this sum is smaller in magnitude than the
%correlation from the  $ns$ state only given by Eq.~(9),
%\begin{equation}
%|B|\ <\ \frac{\sqrt3}2\, |F^0_{ns}\langle
%\psi^r_{ns}|\psi^r_{np}\rangle|\,.
%\end{equation}

%We cannot show this in a general way, but.
%arguments in support of this inequality can be made.
%Assume that in initial state all the atomic electrons
%are moving in the same self-consistent field, thus being described
%by the single-particle wave functions of a certain external field
%Schr\"odinger equation.  We can consider the sum (13), but taken over the complete
%set of states $j$ in the 3-dimensional space.

\section {Closure condition and sum rules}

Using closure we can write simple relations for amplitudes like
$\tilde T_d$ in Eq.~(14), but now  summed over various states $j$.
We will do it in a $3d$ formalism. We will sketch these relations
here and complete their proof subsequently. The closure condition
for the wave functions of the initial state nonrelativistic
Hamiltonian can be written as
\begin{equation}
\sum_{j}|\psi_{j}\rangle \langle \psi_{j}|=1= \sum_{\tilde
j}|\psi_{\tilde j}\rangle \langle \psi_{\tilde
j}|+\sum_{j^*}|\psi_{j^*}\rangle \langle\psi_{j^*}|
+\sum_{j_c}|\psi_{j_c}\rangle \langle \psi_{j_c}|,
\end{equation}
with $\tilde j$ and $j^*$ labelling the occupied and unoccupied
states of discrete spectrum correspondingly, while $j_c$ are
suitably normalized continuum states. The closure condition can be
represented as
$$\int\frac{d^3Q}{(2\pi)^3}\psi^*_{\bfq}(\br)\psi_{\bfq}(\br')+\sum_{n',\ell',m'}
\psi^*_{n',\ell',m'}(\br)\psi_{n',\ell',m'}(\br')=\delta(\br-\br'),$$
for $\Psi_Q$ which are asymptotically plane waves as in Eq.(3).
For any state $j$, occupied or not, we can write, generalizing
Eq.~(15),
\begin{equation}
\langle\psi_f^0|\hat\gamma|\psi_j\rangle=\frac{(\be \cdot \bp)}{m}
\langle\psi_f^0|\psi_j\rangle;  \quad A_j=\frac{(\be \cdot
\bp)}{m} \langle\psi_f^0|\psi_j\rangle\langle\psi_j|\ln
r(1-t)|\psi_i\rangle.
\end{equation}
We may evaluate Eqs. (14),(15) using Eqs.~(16) for the sum
$\sum_{\tilde j}|\psi_{\tilde j}\rangle \langle \psi_{\tilde j}|$ over
occupied bound states.
\begin{equation}
\tilde T_d\ =\ T-T^*_{d}-T_c\,,
\end{equation}
with
\begin{equation}
T \equiv\ \Sigma_jA_j=\langle\psi_f^0|\hat\gamma\ln r(1-t)|\psi_i\rangle
=\frac{(\be \cdot \bp)}{m}\langle\psi_f^0|\ln r(1-t)|\psi_i\rangle
\end{equation}
the sum over the complete set of states $j$, while $\tilde T_{d}$
 and $T^*_d$ are the sums over occupied and unoccupied states
$j$ of the discrete spectrum correspondingly, with $T_c$ the sum over
the continuum states.

We can write Eq. (18) in the form
\begin{equation}
T\ =\ T_d+T_c\,,
\end{equation}
with
\begin{equation}
T_d\ =\ \tilde T_d+T^* ,
\end{equation}
the sum over occupied and unoccupied states of the discrete spectrum.

We may now separate the continuum amplitude $T_c$ defined below
Eq.(19), for which the continuum states $j_c$ may be labeled by
asymptotic momentum $Q$, in two parts
\begin{equation}
T_c\ =\ T_{c1}+T_{c2}\,; \quad T\ =\ T_d+T_{c1}+T_{c2}\,.
\end{equation}
$T_{c1}$ will sum states for which $q \ll P$, $T_{c2}$ states with $Q
\sim P$. More precisely, we pick a $Q_0$ for which $p \gg Q_0 \gg
\tau_c$, with $\tau_c$ the  characteristic momentum of the bound state
(one can assume $\tau_c \approx m\alpha Z$), and define $T_{c1}$ as the
sum over states $Q<Q_0$, $T_{c2}$ as the sum over $Q>Q_0$.

We may show  (see Appendix B) that for states in $T_{c2}$ the wave
function $\psi_Q$ may be replaced by a plane wave:
$\psi_Q^0=\exp{[i(\bf Q \cdot \br)]}$. We may also show (see Sec.~4)
that the sum over plane wave states $\sum_{Q>Q_0}|\psi_{Q}^0\rangle
\langle \psi_{Q}^0|$ in $T_{c2}$ may be extended to a sum over all
plane wave states $\sum_Q$, that is, the sum over plane wave states,
$Q<Q_0$ makes no contribution in $T_{c2}$. But
$\sum_{Q}|\psi_{Q}^0\rangle \langle \psi_{Q}^0|$ over plane wave states
is a sum over a complete set of states (=1), and therefore for $T_{c2}$
one again obtains Eq.~(19), i.e. in the asymptotics
\begin{equation}
T\ =\ T_{c2}\, .
\end{equation}

In Appendix C we show that the amplitude $T_{c2}$ has the same
asymptotics as $T_d$. Hence, while from Eqs. (19), (20) and (21)
\begin{equation}
T\ =\ \tilde T_d+T^*_d+T_{c1}+T_{c2}\,,
\end{equation}
Eq. (23) implies
\begin{equation}
 0=\tilde T_d+T^*_d+T_{c1}; \quad  \tilde T_d=-T^*_d-T_{c1}\,,
\end{equation}
giving an alternative way to calculate $\tilde T_d$ which, as we shall
see, has some advantages.  (In fact all these are asymptotic
amplitudes, and so for Eq.~(25) to follow from (23) and (24) we must
show that these are all amplitudes of the same order. We shall do this
in Sec.~4.) In Sec.~4  we shall make explicit calculations of the
amplitudes of Eq.~(25), determining $\tilde T_d$ from $T^*_d$ and
 $T_{c1}$, as well as further discussing its direct calculations in
Sec.~5.

\section{Formalism for particular amplitudes}.

a In $3d$ formalism all of the particular amplitudes of the
previous section are of the form of Eq.~(14),except that the
summation $\sum_{\tilde j}$ in $\tilde T_d$ is replaced by
summation $\sum_{j^*}$ in $T^*_d$ for unoccupied bound states, and
by integration $\int \frac{d^3Q}{(2\pi)^3}$ in $T_c$ for continuum
states with $Q<Q_0$ and $Q>Q_0$ for $T_{c1}$ and $T_{c2}$
correspondingly. Also, $T$, corresponding to summing over the
complete set of states, was given explicitly in Eq.~(19).

We can also represent the amplitudes in terms of radial functions.
Using Eqs. (2) and (10) we can write
\begin{equation}
T_d\ =\ -\frac{2\sqrt 3\pi^{1/2}\tau}{m}
\frac{(\be \cdot \bp)}{P^4}\sum_{n'}N^r_{n's}
\langle \psi^r_{n's}|\psi^r_{np}\rangle\,
\end{equation}
and correspondingly for partitions $\tilde T_d$ and $T_d^*$
summing over occupied or unoccupied bound states $n's$.

Continuum radial wave functions for $s$ states (as well as those
for a nonzero value of $\ell$) are normalized by condition
$$\int dr \psi^{r*}_{\varepsilon, s}(r)\psi^r_{\varepsilon',s}(r)=\delta(\varepsilon
-\varepsilon'),$$ with
$\psi^r_{\varepsilon',s}(r)=\psi^r_{p,s}(r)/(2\pi p)^{1/2}$,
$\psi^r_{\bp}(\br)=\psi^r_{p,s}(r)/2p+$terms with nonzero values
of $\ell$. The closure relation is
$$\int d\varepsilon \psi^{r*}_{\varepsilon, s}(r)\psi^r_{\varepsilon,s}(r')+\sum_{n',s}
\psi^{r*}_{n',s}(r)\psi_{n',s}(r')=\delta(r-r').$$
 Thus the
calculation of $T_{c1}$ proceeds in the same way again (see
Appendix~A), and yields Eq.~(26), except that for the radial
functions we write $\int_0^\infty d\varepsilon$ instead of
$\sum_{n'}$, and replace $N^r_{n's}$ by $N^r_{\varepsilon s}$.
(The integration over energies has been extended to infinity, with
the energies exceeding strongly the binding energy of the ionized
state, including those for which $\varepsilon >Q_0^2/2m$,
providing a negligible contribution.)

For $T_{c2}$, where $Q>Q_0$, we may argue (Appendix B) that in
$F^0_Q$, corresponding to Eq.~(2), $\psi_Q^0$ is to be replaced by
a plane wave. Thus $F^0_Q=\frac{(\be \cdot \bp)}{m}\delta (\bp-\bf
Q)$. Hence, with plane wave for $\psi_Q$ we may extend the
integration over $Q$ to include $Q <Q_0 \ll P$, since there is no
contribution from this region. On the other hand, we can evaluate
explicitly (see Appendix~C)
\begin{equation}
T_{c2}\ =\ \frac{(\be \cdot \bp)}{m}
\frac{6\sqrt{3\pi}N^r_{np}\tau}{P^4}\,.
\end{equation}
As we have already seen, $T=T_{c2}$. Note that we have also now shown
that all these terms are of the same asymptotic order, which we had
needed to prove Eq.~(25).

Note that we can write Eq.~(26) in the form
\begin{equation}
\sum_{n'}F^{(0)}_{n's}S_{n's,np} +\int\frac{d^3Q}{(2\pi)^3}\,
F^{(0)}_QS_{Q,np}\ =\ 0,
\end{equation}
with $S_{j,i}$ defined by Eq. (5), $F^{(0)}_{n's}$ are the high energy
IPA photoionization amplitudes \cite{A} while $F^{(0)}_Q$ is the
bremsstrahlung amplitude in the  tip region \cite{PT}. Following the previous analysis, the integral
in the second term involves all values of $Q$. However it is saturated by $Q \sim \tau=m\alpha Z$.

We can now write Eq.~(25), expanding the
radial function $\psi^r_{np}(r)$ in terms of the functions
$\psi^r_{xs}(r)$ ($x=n',\varepsilon)$:
\begin{equation}
\psi^r_{np}(r)\ =\
\sum\psi^r_{n's}(r)\,a_{n's,np}+\int\limits_0^\infty
d\varepsilon\psi^r_{\varepsilon s}(r)a_{\varepsilon s,np}\,,
\end{equation}
with
\begin{equation}
a_{n's,np}=\langle \psi^r_{n's}|\psi^r_{np}\rangle\,, \quad
a_{\varepsilon s,np}=\langle\psi^r_{n's}|\psi^r_{np}\rangle\,,
\end{equation}
while closure can be written as
\begin{equation}
\sum_{n'} a^2_{n's,np}+\int\limits_0^\infty d\varepsilon\,
a^2_{\varepsilon ns,np}\ =\ 1.
\end{equation}

(Note that $a_{xs,np}=-2/\sqrt{3}S_{xs,np}$, with $S_{xs,np}$ given by
Eqs.~(6) and (9)).  Note that integrals on the RHS of Eqs. (29) and
(31) are saturated by energies of the order of the $np$ electron
binding energy.

%&&
%T^r =\langle\psi_f|\hat\gamma\ln(1-t)|\psi_i\rangle; \quad
%T^r_d=\sum_{n'}\langle\psi_f|\hat\gamma|\psi^r_{n's}\rangle\cdot
%a_{n'n};
%\nonumber\\
%&& T^r_c\ =\ \int d\varepsilon\langle\psi_f|\hat\gamma|
%\psi^r_{\varepsilon s}\rangle\cdot a_{\varepsilon n}\,.

%Here $a_{xn}(x=n',\varepsilon)$

The ratio of correlations of the $np$ state, with $n's$ and
$\varepsilon s$, to correlation with the $ns$ state may be
described by the factors
 \begin{equation}
x_{n's,np}=\frac{N^r_{n's}a_{n's,np}}{N^r_{ns}a_{ns,np}}\,, \qquad
x_{\varepsilon s,np}=\frac {N^r_{\varepsilon
s}a_{\varepsilon,n}}{N^r_{ns}
 a_{n,n}}\,.
\end{equation}
We can write Eq. (28) in the form
\begin{equation}
\sum_{n'}x_{n's,np}+\int\limits_0^\infty d\varepsilon
x_{\varepsilon s,np}=0\,.
\end{equation}

We must calculate the physical value of the total correlation,
relative to the correlation between $np$ and $ns$ electrons, i.e.
the sum of relative correlations over occupied states $\tilde n$,
\begin{equation} x_{ph}\ =\ \sum_{\tilde n} x_{\tilde ns,np}\,,
\end{equation}
measuring the total amount of correlation relative to the intrashell
$ns,np$ correlation.

\section{Calculation of the physical value \boldmath$x_{ph}$}

Using Eq. (32) for $x_{n's,np}$ one can see that there is a
tendency of cancellation of correlation effects for the $2p$
electrons. Employing Eqs.~(12) and using Eq.~(31) we
find $|a_{n's2p}|\ll1$ for $n'\neq2$.
Since the normalization factors $N^r_{n's}$ drop when $n'$
increases, the contribution of these states to the RHS of Eq.~(34) can
be neglected for $n'>2$. The case of $n'=1$ requires additional
analysis since $N_{1s}>N_{2s}$.  One can see that the contributions of
$n'=1$ and $n'=2$ to $x_{ph}$ have different signs;
from Eq.(12) $a_{2s,2p}<0$, while $a_{1s,2p}>0$, since the
function $\psi^r_{1s}$ has no nodes, while the function
$\psi^r_{2p}$ is non-negative.  Hence, there is
a partial cancellation between these terms. We shall discuss the
extent of this cancellation subsequently.

Now we try to calculate $x_{ph}$ in another way, i.e. as the negative
of the total contribution of low energy unoccupied states, bound
$(x_d)$ and continuum ($x_c$):
\begin{equation}
x_{ph}=\sum_{\tilde n} x_{\tilde n,n}=-x_d-x_c , \quad
x_d=\sum_{n^*} x_{n^*s,np}\,, \quad x_c=\int d\varepsilon
x_{\varepsilon s ,np}
\end{equation}
with $n^*$ labelling the unoccupied states of the discrete spectrum.

\subsection{Discrete states}

%Using Eq.(43) we obtain
%\begin{equation}
%x_{n',n}\ =\ \frac{N_{n's}\langle\psi^r_{n's}|
%\psi^r_{np}\rangle}{N_{ns}\langle \psi^r_{ns}|\psi^r_{np}\rangle}\,,
%\end{equation}
%for all values of $n'$.

It is known that for  $n'\gg n$ the dominant region of coordinate
space in the integral $\langle\psi^r_{n's}|\psi^r_{np}\rangle$ is
determined by the characteristic size of the  $np$ state. These
values of $r$ are much smaller than the characteristic size of the
$n's$ state.  The dependence on the energy of the $n'$-th state in
the Schr\"odinger equation for $\psi^r_{n's}$ can be dropped
\cite{BS}, and the only $n'$ dependence of this matrix element is
contained in the normalization factors $N_{n's}$. Thus, from
Eq.~(32), the ratio
%
%\begin{equation}
%\frac{\langle\psi^r_{n's}\mid\psi^r_{np}\rangle}{N_{n's}}\ \equiv\ g_n
%\end{equation}
%does not depend on $n'$. The factors $g_n$ cancel in the ratios
%
\begin{equation}
x_{n's,np}\ =\ \frac{N_{n's}^2}{N_{ns}^2}\,.
\end{equation}

To estimate $N_{n^*s}^2$ for the unoccupied states with
principal quantum numbers $n^*$ we use the quantum defect approach, in
which the binding energy $E_{n'}$ of the $n's$ state is
\begin{equation}
E_{n'}\ =\ -~\frac\nu{2(n'-\Delta_{n'})^2}\,, \label{56}
\end{equation}
with
$\nu=m\alpha^2$, and
\begin{equation}
\label{57}
\Delta_{n'}\rightarrow\Delta \quad \mbox{ as  }\ n'\to\infty\,.
\end{equation}
Here the quantum defect $\Delta$ does not depend on $n'$.

Combining Eq. (37) with the Fermi--Segre formula \cite{FS},
\begin{equation}
 N_{n's}^2\, =\, 4Z\frac{dE_{n'}}{dn'}\,,
\label{58}
\end{equation}
we obtain
\begin{equation}
x_{n's,np}\ =\ \left(\frac{n-\Delta_n}{n'-\Delta_{n'}}\right)^3
\cdot\kappa_{n',n}\, , \label{59}
\end{equation}
with
\begin{equation}
\kappa_{n',n}\ =\ \frac{1-\Delta'_{n'}}{1-\Delta'_n}\,.
\label{60}
\end{equation}

We neglect the derivatives $\Delta'_{n'}$, as justified below.
Applying Eq.~(39) to the highest occupied state $\tilde n_h$ of
known binding energy, we find its quantum defect $\Delta_h$.
Alternatively, for the lowest unoccupied level of $s$ electrons,
$n^*=\tilde n_h+1$, we identify the quantum defect $\Delta_{n}$
with the limiting value $\Delta$, defined by Eq.~(\ref{57}). The
latter can be extracted from the results of \cite{M}, on the phase
shifts with respect to Coulomb values $\delta(E)$, since
$\Delta=\delta(0)/\pi$. The values of $\Delta_h$, calculated by
using the normalization factors $N^2_{n's}$ given in \cite{HS},
and $\Delta$, are shown in Table~1. We will assume that
$\Delta_{n}$ for all unoccupied states is independent of $n$,
$\Delta_{n}=\Delta$. The comparison of $\Delta_h$ and $\Delta$
indicates the type of error that is being made. Relatively small
value of the difference $\Delta_h-\Delta$ justifies neglect of the
derivatives $\Delta'_{n'}$.

\begin{table}[h]
\begin{center}
\begin{tabular}{|c|c|c|}\hline

$Z$ & $\Delta_h$ & $\Delta$\\ \hline
&&\\
5 & 0.96 & 0.76\\
7 & 1.23 & 0.95\\
10 & 1.44 & 1.27\\
14 & 2.00 & 1.69\\
18 & 2.31 & 2.04\\
32 & 3.04 & 2.74\\
36 & 3.28 & 3.06\\
50 & 3.96 & 3.37\\
\hline
\end{tabular}
\caption{The quantum defects of the highest occupied bound states
$\Delta_h$, and the asymptotic values $\Delta$ as defined by Eq.~(38),
obtained from \cite{M}; $Z$ is the nuclear charge.}
\end{center}
\end{table}

The total contributions  of unoccupied discrete levels
can be written as as
\begin{equation}
x_d\ =\ \sum_{n_h+1} x_{n^*s,np},
\end{equation}
with $n_h$ the principal quantum number of the highest occupied state.
The summation over unoccupied states in Eq.~(42) can be carried out by
using the formula \cite{10},
\begin{equation}
\sum^\infty_{\tilde n_h+1} \frac1{(k+a)^3}\ =\ -\,\frac12\,
\psi''(a) -\sum^{\tilde n_h}_0\,\frac1{(k+a)^3}\,, \label{62}
\end{equation}
with $\psi(a)=\Gamma'(a)/\Gamma(a)$, where $\Gamma(a)$ is the Euler
gamma function, $\tilde n_h+1$ is the principle quantum number of the
lowest unoccupied state.

\subsection{Continuum states}

Equations (37) and (39) reflect the Coulomb-like behavior of the
excited states of the discrete spectrum at $n'\to\infty$. Thus for
the continuum states with $\varepsilon=0$ we can write
\begin{equation}
x_{0s,np}\ =\
\frac{1}{2I_0}\cdot\lim\limits_{n'\to\infty}(n'^3x_{n's,np}),
\label{65}
\end{equation}
with $I_0=m\alpha^2/2=13.6\,$eV, or
\begin{equation}
x_{0s,np}\ =\  \frac{4ZI_0}{N^2_{ns}}\,. \label{66}
\end{equation}
Since the only characteristic energy is the binding energy
$E_{np}<0$ of the ionized state, we may suppose that the integral in the last equality of
 Eq.~(35) is saturated by $\varepsilon\sim|E_{np}|$. If we suppose
that $x_{\varepsilon s,np}=x_{0s,np}$ for $\varepsilon<|E_{np}|$
$x_{\varepsilon s,np}=0$ for $\varepsilon>|E_{np}|$ we find
\begin{equation}
x_c\ \approx\ \frac{4ZI_0|E_{np}|}{N^2_{ns}}\,.
\label{67}
\end{equation}
This is clearly a fairly crude estimate, but we will see that it is
consistent with results from direct calculations of $x_{ph}$.

\subsection{Results for \boldmath $x_{ph}$}

Note that in the quantum defect approach $x_d>0$. The assumption (46)
provides also $x_c>0$.
Hence the values $x_{ph}$ are negative. This means that the total
correlation effect in the amplitude has a sign, which is opposite to
that of correlation inside the same shell. Employing Eqs. (42), (43)
and (46) we find the values  of $x_{ph}$  shown in Table~2.

\begin{table}[h]
\begin{center}
\begin{tabular}{|c|c|c|c|c|c|c|}\hline

$Z$ & $n$ & $x_d$ & $x_c$ & $x_{ph}$ & $x^{dir}$\\ \hline
 &  &  &  &  &\\

5 & 2 & 0.08 & 0.12 & -0.20&     \\
7 & 2  & 0.05 & 0.09 & -0.14& -0.18\\
10 & 2 & 0.04 & 0.07 & -0.11& -0.11\\
14 & 2 & 0.06 & 0.16 & -0.22&      \\
14 & 3 & 0.08 & 0.12 & -0.20&      \\
18 & 2 & 0.06 & 0.20 & -0.26&      \\
18 & 3 & 0.05 & 0.10 & -0.15&   -0.14  \\
32 & 3 & 0.01 & 0.18 & -0.19&     \\
32 & 4 & 0.13 & 0.11 & -0.24&    \\
36 & 3 & 0.01 & 0.19 & -0.20&    \\
36 & 4 & 0.08 & 0.11 & -0.19&    \\
50 & 5 & 0.06 & 0.13 & -0.19&    \\
\hline
\end{tabular}
\caption{The values of $x_d$, $x_c$ and $x_{ph}$ as defined by
Eq.~(35), where $n$ is the principal quantum number of the ionized $np$
state, $Z$ stands for nuclear charge.  The values $x^{dir}$ presented
in the last column are the results of direct summation over occupied
states -- Eq.~(34).}
\end{center}
\end{table}

Our results are in good agreement with the results $x^{dir}$
obtained by direct
 summation of the correlations with occupied shells in the
 photoionization amplitude \cite{5}. Results of \cite{5}
 obtained by inclusion of correlations with $2s$ and $1s$ shells in
 nitrogen and neon are $x_{ph}=-0.18$ and $x_{ph}=-0.11$
 respectively. Since the results of \cite{5} are in good agreement
 with those of \cite{4} for angular distributions, our results agree
with those of \cite{4} as well. For ionization of the $3p$ state in
argon, ``shell by shell" calculation \cite{5} gives $x_{ph}=-0.14$,
also in agreement with the result of the present work.

Now we estimate the total contribution of correlations to the
amplitudes and cross sections of photoionization of $p$ states. We use
the estimate $\langle\psi_{np}|\psi_{ns}\rangle\approx-1$ for all $n$
(see Eq.~(12)). Presenting the ratio of IPA amplitudes in terms of
normalization factors $F_{n,0,0}/F_{n,1,0}=N^r_s\big/\sqrt3\,N^r_p$, we
find
\begin{equation}
F_{n,1,0}\ =\
F^0_{n,1,0}\left(1+\frac{N^r_{ns}}{2N^r_{np}}\,x_{ph}\right),
\end{equation}
and thus the cross section for ionization of $np$ state beyond IPA
is
\begin{equation}
\sigma_{np}\ =\ \sigma^0_{np}\left(1+\frac{N^r_{ns}}{3N^r_{np}}\,
x_{ph}\right),
\end{equation}
with $\sigma^0_{np}$ standing for the IPA values.

Using the numerical values of the normalization factors \cite{HS}
we find that the total correlations diminish the values of cross
sections of photoionization of $2p$ states in nitrogen and neon
only by about 2.5\%. In contrast, inclusion of correlations only
with the $2s$ shell would increase the cross sections by 18\% and
22\% correspondingly.The full cross section for ionization of $3p$
states in argon becomes smaller by 1.8\%, while it becomes larger
by 12\% if only correlation with the $3s$ shell is considered.

\section{Calculation for hydrogenlike functions \newline and
limitations for the many-electron atoms}

We can also make explicit calculations of correlations in the case of
Coulomb wave functions, and also in using an effective charge
approximation for screening.  All results for Coulomb functions can be
obtained analytically.  The results for $x_{n's,np}$ do not depend on
the values of nuclear charge $Z$.

Starting with ionization of  $2p$ electrons we obtain
the values of the parameters that are presented in Table~3 (upper indices
$C$ indicate that the quantities are calculated in the Coulomb
field of the nucleus).

\begin{table}[h]
\begin{center}
\begin{tabular}{|c|c|c|}
\hline
$n'$ & $x^C_{n's,2p}$ &$ x^C_{n's,3p} $\\
\hline
1 &  -1.58&-1.26 \\
2 & ~1.00& -0.02\\
3 & ~0.041&1.00\\
4 & ~~0.015&0.04\\
\hline
\end{tabular}
\caption{Parameters for ionization of $2p$ and $3p$ states,
obtained by using Coulomb functions.}
\end{center}
\end{table}

One can see that in this case the correlation with the $1s$ state
is about 50\% larger than that with the $2s$ state.  For the
contribution of the continuum with $\varepsilon\ll E$ one can
obtain by direct calculation
\begin{equation}
x_{\varepsilon2p}=\frac{C\Phi(\varepsilon)}{(\varepsilon+I_Z/4)^2};
\quad \Phi(\varepsilon)=\exp(-2\xi_1(\arctan(2/\xi_1))-2/\xi_1)), \quad
\xi_1=\tau/\varepsilon=\sqrt{I_Z/\varepsilon}; \quad \Phi(0)=1,
\end{equation}
where
\begin{equation}
C\ =\ 2\lim_{n'\to\infty} n'^3 x^C_{n's,2p}\ =\ 0.78
\end{equation}
is obtained by using the well known Coulomb wave function
for the bound $n's$ state for $n'\gg1$ \cite{BS}. This gives
$x_c=0.52$, in
agreement with Eq.~(33).

%Note that closure provides $U/U(2)=1/3$ in the
%Coulomb case.

Consider now ionization of $3p$ electrons - see again Table~3. The
correlation with $1s$  and $3s$ states cancel to a larger extent
than in the case of the $2s$ state. The contribution of the
continuum is now $x_c=0.22$.

% while $U/U(3)=1/3$.

Now we analyze the situation for more realistic atomic models. We
shall compare the values $x_{ph}$ calculated is the unscreened and
screened Coulomb fields. Here we obtain $x_{ph}$ as the sum over
occupied states, taking into account that, as we showed above, for
ionization of a $np$ state only correlations with $ns$ and $1s$
states are important. Using Eq.(32) we can write in this
approximation
\begin{equation}
x_{ph}=1+x_{1s,np},
\end{equation}
with $x_{1s,np}<0$, as shown above. As one can see from Table 3,
the Coulomb values  $|x^C_{1s,np}|>1$. Now we show that for the
screened Coulomb values $|x_{1s,np}|<|x^C_{1s,np}|$, and thus the
Coulomb values $x^C$ can be used as the lower limits for the
physical values $x_{ph}$.

We can calculate the screening effects, assuming that the initial
electrons are described by the Coulomb functions with effective
values of the nuclear charge $Z_{n \ell}=Z-\delta_{n\ell}$ \cite{
BS,G}. In this approach we find for ionization of $2p$ electrons
\begin{equation}
x_{1s,np}=x^C_{1s,np}\eta; \quad
\eta=\left(\frac{Z_{1s}}{Z_{2s}}\right)^3\left(\frac{3Z}{2Z_{1s}+Z_{2p}}\right)^4
\left(\frac{Z_{2s}+Z_{2p}}{2Z}\right)^5\frac{2Z}{3Z_{2s}-Z_{2p}},
\end{equation}
with $\eta=1$ if screening is neglected and thus
$Z_{1s}=Z_{2s}=Z_{2p}=Z$. In the lowest order of expansion in
powers of $\delta_{n\ell}$ Eq. (52) provides $\eta=1+\delta/Z$,
with $\delta=-\delta_{1s}/3-5\delta_{2p}/3+2\delta_{2s}$. If a
small influence of the electrons in the higher states on the
values of $\delta_{n\ell}$ is neglected, $\delta$ is the same for
all atoms with the totally occupied $K$ and $L$ shells. Using the
values $\delta_{1s}=0.35$, $\delta_{2s}=3.25$ and
$\delta_{2p}=4.75$, \cite{G}, we find $\delta <0$. Thus $\eta<1$,
and indeed
\begin{equation}
x_{phys}>x^C,
\end{equation}
while $x^C<0$. Using Eq.(52) for neon with these values of
$\delta_{n\ell}$ we find $\eta=0.676$, providing
$x_{1s,2p}=-1.068$ and thus $x_{phys}=-0.068$, with $|x_{phys}|$
smaller then that shown in Table 2. Note however that this value
is a result of subtraction of two much larger values. Putting
$\eta=0.702$, i.e. increasing it by $4\%$ we would find
$x_{phys}=-0.11$, in agreement with the data in Table 2.

%\section{Final formulas}

%Now we shall compare the values of correlations inside  $n$ shell
%described by the value $A_{ns}$ presented by
%Eq.~(7) and the value of $T$ defined by Eq.~(12) obtained by the closure
%condition. The ratio
%\begin{equation}
%x\ =\ \frac T{A_{ns}}
%\label{36}
%\end{equation}
%can be evaluated by using
%Eqs. (7), (16), (19) and (\ref{24})
%\begin{equation} x\ =-\
% \frac{N_{np}}{N_{ns}m\alpha Z}\ \frac1{\langle\psi^r_{ns}\mid
% \psi^r_{np}\rangle}\,.
% \end{equation}
% The inequality
% \begin{equation} x\ <\ 1
%\end{equation}
%is equivalent to (14).

%Of course, the values of $x$ can be computed by straightforward
%employing of numerical atomic wave functions. However it is instructive
%to present the ratio of normalization factors $N_{k}$ ($k=ns, np$) in
%terms of their values in the Coulomb field $N^C_{k}$:
%\begin{equation}
%R_n\ =\ \frac{N_{np}}{N_{ns}m\alpha Z}\ =\ r_n\kappa_n
%\end{equation}
%with
%\begin{equation}
%r_n=\frac{N_{np}/N^C_{np}}{N_{ns}/N^C_{ns}}\,; \quad
%\kappa_n=\frac{N^C_{np}}{N^C_{ns}m\alpha Z}\,.
%\end{equation}
%The Coulomb values $\kappa_{n}$ do not depend on $Z$, exhibiting very
%weak dependence on the quantum number $n$ \cite{7}
%\begin{equation}
%\kappa_2=\frac{\sqrt3}6\approx0.29; \quad
%\kappa_3=\frac{2\sqrt2}9\approx 0.31; \quad
%\kappa_4=\frac{\sqrt{5/3}}4\approx 0.32\,.
% \end{equation}
%Thus, for the estimations we can assume
%\begin{equation}

\section{Exclusive and inclusive processes in neutral atoms and ions}

Instead of photoionization of neutral atoms we can consider
photoionization of ions. Since the cancellation of correlations is
due mainly to cancellation  between correlations with $1s$ and
with $ns$ electrons, a hole in either the $1s$ or $ns$ shell
breaks this balance, and the net correlations will greatly
increase.  This was observed earlier \cite{3} for the cases of
nitrogen and neon. A hole in other shells will not influence
strongly the total correlation.

One is often interested in inclusive and exclusive
photoionization, as in ionization of a $2p$ state, but perhaps
also exciting  other electrons. The theory of such processes was
much studied \cite{S1}--\cite{S3} in the case of shake off and
shake up.

As already noted, our discussion considering only two active
electrons, was inclusive in its treatment of the spectator
electrons, it could be exclusive if overlap integrals between
initial and final spectator states were considered. Note that in
principle one should have also, in the presence of correlations,
include other electrons as active, capable to undergo further
excitation or ionization beyond the shakeoff/shakeup mechanism. We
can try to estimate the magnitude of these various mechanisms.

Assume that the bound electrons, moving in a certain
self-consistent field , find themselves in another field after the
electron is ejected. Photoionization of an $n, \ell$ state can be
followed by a transition of an electron $n',\ell'$ to the state
$n^*,\ell^*$.
 One should consider the
process simultaneously with ionization of $n',\ell'$ state
followed by a transition of an electron $n,\ell$ to the state
$n^*,\ell^*$. The asymptotic IPA amplitude of the process, without
correlations, is
\begin{equation}
{\cal F}^0= F^0_{n \ell}\langle\phi_{n^*\ell^*}
|\psi_{n'\ell'}\rangle-\ F^0_{n' \ell'}\langle\phi_{n^*,\ell^*}
|\psi_{n \ell}\rangle,
\end{equation}
with $\psi$ and $\phi$ the functions in the fields of the atom and
of the ion with a hole in ${n, \ell}$ and ${n',\ell'}$ states for
the two terms in the right hand sides (RHS) of Eq.(54)
correspondingly, $F^0_{n\ell}$ is the asymptotic IPA amplitude for
ionization of $n\ell$ state. One needs $ \ell^*=\ell'$, or
$\ell^*=\ell$ otherwise the matrix element vanishes due to
orthogonality of the angular parts of the wave functions.

Correlations provide another mechanism of the process in which $n,
\ell$ electron is ionized by direct interaction with the photon,
and in the next step the photoelectron excites the $n'\ell'$
electron to $n^* \ell^*$ state. One should include possible
permutation of the $n\ell$ and $n'\ell'$ states. The amplitude,
which includes the correlations can be written as

\begin{equation}
{\cal F} ={\cal F}^0+ F^0_{n
\ell}\Lambda_{n^*\ell^*,n'\ell'}-F^0_{n'
\ell'}\Lambda_{n^*\ell^*,n\ell},
\end{equation}
with ${\cal F}^0$ given by Eq.(54). The correlations can cause
these transitions even if $\ell^*$ coincides neither with $\ell$
nor with $\ell'$, and the shakeup mechanism can not contribute.

Note that the contribution of correlations on the RHS of Eq.(55)
is written  omitting the terms containing as additional factors
the overlap matrix elements of the type $\langle\phi_{n' \ell}
|\psi_{n' \ell}\rangle$. We neglected such terms in particular
calculations through the paper-see Sec.3. Inclusion of such terms
would not alter the asymptotic energy dependence of the amplitude.
Hence we shall use Eq.(55).

Consider, for example, photoionization of Be with the final state
ion containing electron excited into $2p$ state. In this case
$n=1, n'=n^*=2, \ell=\ell'=0, \ell^*=1$. Both terms on the RHS of
Eq.(55) turn to zero, providing ${\cal F}^0=0$, and thus
\begin{equation}
{\cal F} = F^0_{1s}\Lambda_{2p,2s}-F^0_{2s}\Lambda_{2p,1s}.
\end{equation}

Now we study the the relative role of the shakeup and correlation
mechanisms of the process, for various relations between $\ell$,
$\ell'$ and $\ell^*$. To obtain the energy dependence of the
contributions on the RHS of  Eq.(55) one can employ that in the
asymptotics
\begin{equation}
F^0_{n \ell}\sim\omega^{-(3+\ell)/2} \quad \Lambda_{n*\ell*,n\ell}
\sim \omega^{-1/2}.
\end{equation}
The estimation  for $F^0_{n \ell}$ is well known \cite{BS}. The
estimation for  $\Lambda_{n*\ell*,n\ell}$ is the consequence of
Eq.(4), in which the matrix element of $\Lambda$ between the bound
states is proportional to the factor $\xi \sim \omega^{-1/2}$.

As we have seen for $\ell \neq \ell' \neq \ell^*$ only
correlations contribute. Turn now to other cases. If $\ell \neq
\ell'=\ell^*$ the second term on the RHS of Eq. (54) vanishes and
the IPA amplitude is
$$ {\cal F}^0=F^0_{n \ell}\langle\phi_{n^* \ell'}|\psi_{n' \ell'}\rangle.$$
Using Eq.(57) we find that the second term on the RHS of Eq.(55)
drops faster than ${\cal F}^0$, and hence the asymptotics is
\begin{equation}
{\cal F}= F^0_{n \ell}\langle\phi_{n^* \ell'} |\psi_{n'
\ell'}\rangle-F^0_{n' \ell'}\Lambda_{n*\ell',n\ell}.
\end{equation}
Further analysis depends on relation between $\ell$ and $\ell'$.

For $\ell'=\ell^*<\ell-1$(for example, $n's \rightarrow n^*s$ and
$n'p \rightarrow n^*p$ transitions in ionization of $d$ states),
the correlations determine the asymptotics of the process since
the first term on the RHS of Eq. (58) drops with energy faster
then the second term. For $\ell'=\ell^*=\ell-1$ (for example, $n's
\rightarrow n^*s$ transitions in ionization of $p$ states) states
the two terms terms on the RHS of Eq.(58) behave with energy in
the same way. However the first term is proportional to the
overlap matrix element, which is usually small. If it is the case,
the correlations dominate the process.

Similar analysis shows that for $\ell'=\ell^*>\ell-1$, including
the case $\ell=\ell'=\ell^*$ (for example, $n's \rightarrow n^*s$
transitions in ionization of $s$ states) asymptotics is determined
by the shakeup mechanism, described by the first term on the RHS
of Eq.(58). However, at finite energies, where the experimental
data is available, interplay of the shakeup and correlation terms
appears to be important \cite{DA}, \cite{PS}.

To obtain the cross section for the inclusive process one should
sum the squared amplitude given by Eq.(55) over $n'$ and $n^*$ ,
depending on the conditions of
 experiment.

\section{Summary}

We have calculated the IPA breaking correlation corrections to the
 high energy photoionization amplitude, focusing on ionization of $p$
states. Instead of carrying out summation over occupied states, we
 employed the closure results for summation over all states of the
 spectrum. We showed that the sum over all states of the spectrum is
 equal to the contribution of its high energy part. Therefore there is
 total cancellation between contributions of discrete and low energy
part of continuum spectra.  This provided identities, involving the
asymptotics of the amplitudes of photoionization and of bremsstrahlung
amplitudes at the tip region -- Eqs. (28), (33).

We calculated the sum of correlations with the occupied bound
states as the negative of the sum over the unoccupied bound states and
the low energy continuum states.  We made conclusions in a simple
model, based on the general features of the bound states with large
 principal quantum numbers $n$. In this approach the sum of the
 occupied state correlations has a sign, which is opposite to that of
 the correlation inside the same shell. In spite of a crude model for
 the continuum, the results are in good agreement  with those obtained
 earlier in direct calculation (Table~2).

 We have shown that there is a general tendency of cancellation for the
 correlation effects. We demonstrated also that calculations with
 Coulomb functions give limits for the correlation
 effects in screened atoms, Eq.(53).

We showed also that the correlations beyond shakeoff and shakeup
 effects are important in inclusive processes, where photoionization is
accompanied by excitation of other electrons. The relative role of
the shakeup and correlation mechanisms was found to depend on
relations between orbital momenta $\ell$ and $\ell'$ of the
removed electrons and orbital momentum $\ell^*$ of the excited
electron. In some of the cases the correlations dominate in the
process. In the particular cases, for which experimental data are
available , interplay of the two mechanisms is important.

\subsection*{Acknowledgment}
We thank S. T. Manson for fruitful discussions. One of us (EGD)
wishes to express his thanks for hospitality during his visits to
the University of Pittsburgh. The work was supported in part by
NSF grant 0456499.

\def\thesection{Appendix \Alph{section}}
\def\theequation{\Alph{section}.\arabic{equation}}
\setcounter{section}{0}

\section{}
\setcounter{equation}{0} We calculate the asymptotic amplitude (2)
for the bound ${n's}$ states, following the approach of \cite{A}.
Note that it is determined at small $r\sim1/P$. Thus we can use
expansion of the function $\psi_{ns}(r)$ at $r\to0$,
\begin{equation}
\psi_{ns}(r)\ =\ \frac{N^r_{ns}}{\sqrt{4\pi}}(1+a_sr)\exp(-\lambda r);
 \quad \lambda>0, \quad \lambda\to0.
\end{equation}
Here the last factor has been introduced to insure the convergence
of the integral in the intermediate steps. We have kept two terms
of expansion in powers of $r$ in brackets, since the lowest one,
as we shall see below, vanishes at $\lambda=0$. (Higher terms in
$r$ would contribute in higher terms in $1/P$). The parameter
$a_s$ on the RHS of Eq.~(A.1) should is equal to the first
derivative of the function $\psi_{ns}(r)$, as determined by the
first Kato cusp condition \cite{11}, being $a_s=-m\alpha Z=-\tau$.
Since
\begin{equation}
\int d^3re^{i(\bp \cdot \br)-\lambda r}=\left(
-\frac\partial{\partial\lambda}\right) \int d^3re^{i(\bp \cdot
\br)}\,\frac{e^{-\lambda r}}r=
\frac{8\pi\lambda}{(P^2+\lambda^2)^2},
\end{equation}
and $re^{-\lambda r}=\left(-\frac\partial{\partial\lambda}\right)
e^{-\lambda r}$, we obtain Eq.~(2).

Evaluation of the  matrix element in Eq.(2) corresponding to a
continuum state {j} with asymptotic momentum $Q \ll P$ can be done
in the same way, with the same form of expansion of $\psi_{Qs}$.
Since only $s$ waves contribute, we can write
\begin{equation}
\langle\psi_f^{(0)}| \hat\gamma|\psi_{Qs}\rangle\ =\ F_{Qs}^0\ =\
\frac{(\be \cdot \bp)}m \,X_{\varepsilon s}\,,
\end{equation}
where $\varepsilon$ is the energy of the continuum electron,
\begin{equation}
X_{\varepsilon s}\ =\ \frac{4\pi^{1/2}N^r_{\varepsilon
s}\tau}{P^4}\,,
\end{equation}
with $N^r_{\varepsilon s}=\psi^r_{\varepsilon s}(0)$, where the
upper index $r$ again denotes the radial part of the function,
yielding Eq.(2).

\section{} % {Appendix B}
\setcounter{equation}{0}

Now we evaluate the amplitude $T_{c2}$ defined, following
Eq.~(17), as
\begin{equation}
T_{c2}\ =\ \frac{(\be\bp)}m \int
\frac{d^3Q}{(2\pi)^3}\,\langle\psi^0_f|\psi_Q\rangle\,X(Q)\,,
\end{equation}
with integration over $Q\gg\tau$, and with
\begin{equation}
X(Q)\ =\ \langle\psi_Q|\ln r(1-t)|\psi_i\rangle\,.
\end{equation}
Using Eq. (9.6) of \cite{BS}, and taking the first iteration, we
obtain
\begin{equation}
\langle\psi^0_f|\psi_Q\rangle\ =\ (2\pi)^3\delta(\bp-\bfq)+h(Q)
\end{equation}
with
\begin{equation}
h(Q)\ =\ 2m\frac{V(\bp-\bfq)}{(Q+P)(Q-P+i\nu)}\, \quad \nu >0,
\quad \nu \to0.
\end{equation}
The first term on the right hand side of Eq.~(B.3) immediately gives
\begin{equation}
T_{c2}\ =\frac{(\be\bp)}{m}X(P)\,,
\end{equation}
leading to Eq. (23) due to Eq.~(19).

Now we show that the second term on the right hand side of
Eq.~(B.3) provides higher order terms of expansion in powers of
$P^{-1}$.

We put $2mV(q)=\tau_cv(q^2)$, with $\tau_c$ being of the order
$\tau\ll P$. One can see immediately that the regions $|\bfq-\bp|
\sim P$ lead to corrections of the order $1/P$. The vicinity of
the point $\bfq=\bp$ requires special analysis. Near this point we
use the well known relation
\begin{equation}
\frac{1}{x+i\nu}\ =\ P.V.\,\frac1x- i \pi\delta(0)\,.
\end{equation}
For the first term on the right hand side of Eq.(B6) the result of
integration over the angles leads to a function of $(P-Q)^2$, i.e.
to an even function of $P-Q$. Together with the denominator $P-Q$
this leads to an odd function of $P-Q$ in the integrand of the
integral over $Q$, providing contribution of the order $\sim
\tau_cX(p)/P$. Contribution of the same order comes from the whole
interval $Q\sim P$. In a similar way one can see that the second
term on the right hand side of Eq.(B6) also contributes only
beyond the asymptotics. Thus, indeed we can neglect the second
term on the right hand side of (B.3).

\section{} %{Appendix C}
\setcounter{equation}{0}

 In order to calculate $T_{c2}$ we must evaluate the matrix element defined by Eq.(35).
It is expressed also by Eq.(30 of appendix B. Since $Q \gg \tau$
we first describe $\psi_Q$ by a plane wave (3). Then
\begin{equation}
X(Q)\ =\ \int dVe^{-iQz} \ln(r-z)\,\psi_i(r)\,.
\end{equation}

For $p$ states we can write
\begin{equation}
\psi_i(r)\ =\ \sqrt{\frac3{4\pi}}\, t\psi^r_i(r)
\end{equation}
(recall that we need only the states with $\ell_z=0$). Since we shall
 need $\psi_i(r)$ at $r\sim1/Q$, we can put
$\psi^r_i(r)=N^r_ire^{-\lambda r}$ $(\lambda\to0)$ in Eq.~(A.4), and
thus $\psi_i(r)=\sqrt{3/4\pi}\,N^r_i re^{-\lambda r}$. We use
$ze^{-iQz}=iD_Qe^{-iQz}$ (with $D_p=\partial/\partial p$) and use
the parabolic coordinates
$$
\xi=r+z, \quad \eta=r-z, \quad \phi=\arctan(x/y),
$$
so that
$$
r=\frac{\xi+\eta}2; \quad z=\frac{\xi-\eta}2,
$$
and
$$
dV\ =\ \frac{\xi+\eta}4\, d\xi d\eta d\phi\,.
$$
Thus we find
\begin{equation}
X(Q)\ =\ \frac{-i\sqrt{3}\, \pi^{1/2}N_{np}}2\, D_QD_\lambda J_Q
\end{equation}
with
\begin{equation}
J_Q=\int d\xi\exp\left(-\frac{\lambda+iQ}2 \right)\int d\eta
\exp\left(-\eta\frac{\lambda-iQ}2\right)
\ln\eta=\frac{-4}{\lambda^2+Q^2}\left(\gamma_E+\ln\frac{\lambda-iQ}2
\right),
\end{equation}
with $\gamma_E \approx 0.578$ the Euler constant, leading to
asymptotics $1/Q^4$.

Note that the asymptotics $Q^{-4}$ of the function $X(Q)$ is due
to the logarithmic term in the integrand. Replacing it by a
constant, we would obtain the asymptotics $Q^{-5}$. Indeed,
replacing $\ln\eta$ by a constant, we would immediately find
$X(Q)=0$ due to the operator $D_\lambda$. Thus we must include the
next term of expansion of the function $\psi_i(r)$ in powers of
$r$ in a way similar to Appendix~A. This gives $X(Q)\sim
D_QD^2_\lambda\cdot 1/(\lambda^2+Q^2)=4/Q^5$. It can be shown that
the higher order corrections to the function $\psi_Q$ contribute
only to higher orders of expansion in powers of $1/Q$.

%\newpage

%\newpage


\begin{thebibliography}{**}


\bibitem{1} E. W. B. Dias {\em et al.}, Phys. Rev. Lett {\bf78}, 4553
(1997).

\bibitem{2} D. L. Hansen, {\em et al}., Phys.  Rev. A~{\bf60}, R2641 (1999).

\bibitem{3} V. K. Dolmatov, A. S. Baltenkov, and S. T. Manson,{\bf64}, 042718
(2001).

\bibitem{4} E. G. Drukarev and  N. B. Avdonina, J. Phys. B~{\bf36},
2033 (2003).


\bibitem {5} E. G. Drukarev and R. H. Pratt, Phys. Rev. A~{\bf72},
062701 (2005).

\bibitem {6} E. G. Drukarev and M. I. Strikman, Phys. Lett B~{\bf 186},
1 (1987).
\bibitem{MN} O. Hemmers, {\em et al}., J. Phys. B {\bf30}, L727 (1997).
\bibitem{DT} E. G. Drukarev and M. B. Trzhaskovskaya, J. Phys. B
{\bf 31}, 427 (1998).

\bibitem{DA} E. G. Drukarev, E. Z. Liverts, M. Ya. Amusia, R. Krivec and V. B. Mandelzweig,
Phys.Rev. A{\bf 77}, 012715 (2008).



\bibitem{BS} H. Bethe and E. E. Salpeter, {\em Quantum Mechanics of One-
and Two-Electron \\ Atoms} (Springer-Verlag, Berlin, 1957).


\bibitem{LL} L. D. Landau and E. M. Lifshitz, {\em Quantum
Mechanics. Nonrelativistic Theory} (Pergamon, Oxford, 1991).

\bibitem{JC} K. T. Cheng and W. R. Johnson, Phys. Rev. A {\bf15},1326 (1977);
{\bf 16}, 263 (1977).

\bibitem{A} T. Aberg, Phys. Rev. A {\bf2}, 1726 (1970).
\bibitem{PT} H. K. Tseng and R. H. Pratt, Phys.  Rev. A~{\bf3}, 100 (1971).

\bibitem{FS} E. Fermi and T. Segre, Z. Phys. {\bf 82}, 729 (1932).

\bibitem{M} S. T. Manson, Phys. Rev. {\bf182}, 97 (1969).

\bibitem{HS} F. Herman, and S. Skillman, {\em Atomic Structure
Calculations}, (Englewood Cliffs, NJ, 1963).

\bibitem{G} J. Burns, J. Chem. Phys. {\bf41}, 1521 (1964).

\bibitem{10} A. Kratzer and W. Franz, {\em Transzendente Funktionen},
(Acad. Verl. Leipzig, 1960).

\bibitem{S1} T. A. Carlson and C.W. Nestor Jr, Phys. Rev. A~{\bf8},
2887 (1973).

\bibitem{S2} A. G. Kochur, A. I. Didenko and D. Petrini, J. Phys. B~{\bf35},
395 (2002).

\bibitem{S3} A. G. Kochur and A. I. Didenko, Optics and Spectroscopy
{\bf100}, 645 (2006).
\bibitem{PS} T. Suric and R. H. Pratt, J. Phys. B {\bf 37}, L93
(2004).

\bibitem{11} T. Kato, Com. Pure Appl. Math. {\bf10}, 151 (1957).


\end{thebibliography}
\end{document}